\documentclass{osa-article}

\journal{osac}

\articletype{Research Article}

\usepackage{lineno}
\usepackage{amsmath}
\usepackage{graphicx}
\usepackage{bm}
\usepackage{appendix}
\usepackage{caption}
\usepackage{multirow}
\usepackage{comment}

\begin{document}

\title{Focal-plane wavefront sensing with photonic lanterns I: theoretical framework}

\author{Jonathan Lin,\authormark{1,*} Michael P. Fitzgerald,\authormark{1} Yinzi Xin,\authormark{2} Olivier Guyon,\authormark{3} Sergio Leon-Saval,\authormark{4} Barnaby Norris,\authormark{5} Nemanja Jovanovic\authormark{3}}

\address{\authormark{1} Physics \& Astronomy Department, University of California, Los Angeles (UCLA), 475 Portola Plaza, Los Angeles 90095, USA\\
\email{\authormark{*}jon880@astro.ucla.edu} 
\authormark{2} Department of Astronomy and Steward Observatory, The University of Arizona, 933 N. Cherry Ave., Tucson, AZ 85719, USA\\
\authormark{3} Department of Astronomy, California Institute of Technology, Pasadena, CA, 91125, USA\\
\authormark{4} Sydney Astrophotonic Instrumentation Laboratory, School of Physics, The University of Sydney, Sydney, NSW 2006, Australia \\
\authormark{5} Sydney Institute for Astronomy, School of Physics, Physics Road, The University of Sydney, NSW 2006, Australia}




\begin{abstract}
The photonic lantern (PL) is a tapered waveguide that can efficiently couple light into multiple single-mode optical fibers. Such devices are currently being considered for a number of tasks, including the coupling of telescopes and high-resolution, fiber-fed spectrometers, coherent detection, nulling interferometry, and vortex-fiber nulling (VFN). In conjunction with these use cases, PLs can simultaneously perform low-order focal-plane wavefront sensing. In this work, we provide a mathematical framework for the analysis of the photonic lantern wavefront sensor (PLWFS), deriving linear and higher-order reconstruction models as well as metrics through which sensing performance --- both in the linear and nonlinear regimes --- can be quantified. This framework can be extended to account for additional optics such as beam-shaping optics and vortex masks, and is generalizable to other wavefront sensing architectures. Lastly, we provide initial numerical verification of our mathematical models, by simulating a 6-port PLWFS. In a companion paper \cite{paper2}, we provide a more comprehensive numerical characterization of few-port PLWFSs, and consider how the sensing properties of these devices can be controlled and optimized.
\end{abstract}

\section{Introduction}

\begin{figure}
    \centering
    \includegraphics[width=\textwidth]{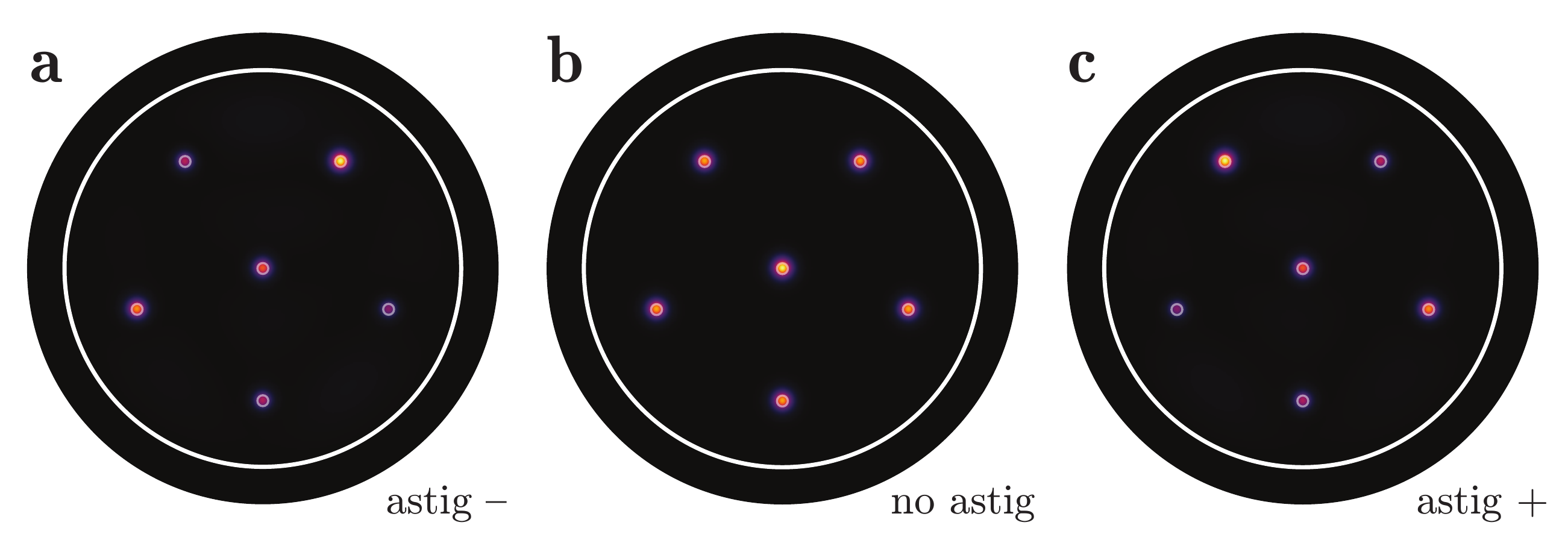}
    \caption{Simulated response of a 6 port lantern in the presence a: -1 rad rms astigmatism; b: 0 rad rms astigmatism; and c: +1 rad rms astigmatism. The photonic lantern converts phase variations into unique intensity variations among the output cores. Circles show the jacket-cladding interface and the cladding-core interfaces. Optical propagation is simulated using the Python packages HCIPy and Lightbeam. }
    \label{fig:astig}
\end{figure}

High-contrast imaging is becoming one of the primary tools for the direct detection and characterization of exoplanets. This class of techniques combines ground-based extreme adaptive optics (AO), which corrects for wavefront aberrations induced by passage of light through the atmosphere and the instrument, and coronagraphy, which suppresses on-axis starlight to reveal the circumstellar environment, as well as contrast-boosting post-processing techniques such as angular differential imaging \cite{Marois:06:ADI} and spectral differential imaging \cite{Sparks:02:SDI}. Together, these techniques enable contrasts down to $\sim 10^{-6}$ and angular separations down to 200 mas. So far, some 30 exoplanets have been detected through high-contrast imaging techniques \cite{Bowler:16}; however, almost all are widely separated gas giants with masses several times that of Jupiter. One of the main roadblocks in increasing current sensitivity are non-common-path aberrations (NCPAs): quasi-static aberrations evolving on the timescale of minutes to hours that occur due to instrument instabilities induced by humidity, temperature, and gravity vector changes \cite{Martinez:12:NCPA1,Martinez:13:NCPA2}. Because these aberrations appear downstream from the wavefront sensor, they cannot be removed via typical pupil-plane wavefront control systems. As a result, wavefront control must be improved before instruments can attain the necessary contrasts and angular separations typical for systems similar to the Sun and Earth: $\sim 10^{-10}$ and $\sim 100$ mas, at a distance of 10 pc, in visible light \cite{Traub:10}. One way forward is to sense wavefront aberrations in the final focal plane with the science camera, so that sensor and science light travel down the same optical path. This approach, known as focal-plane wavefront sensing (FPWFS), removes NCPAs.
\begin{figure}
    \centering
    \includegraphics[width=\textwidth]{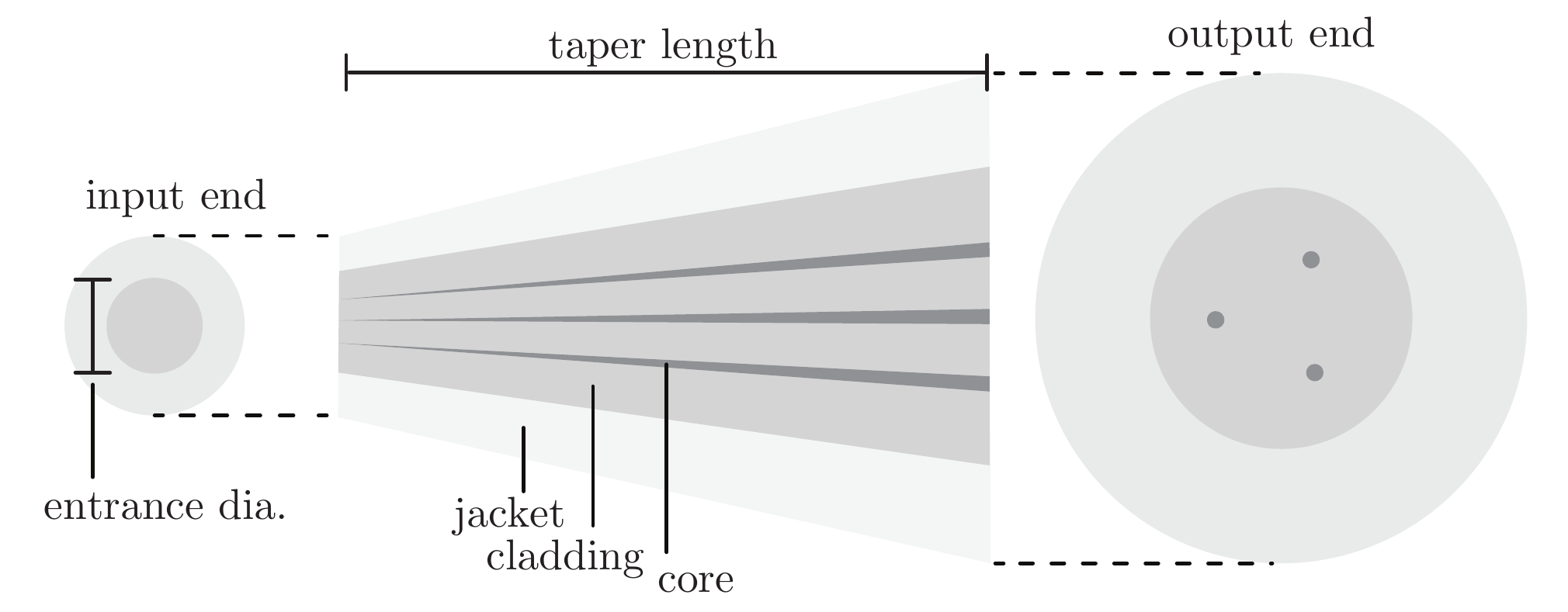}
    \caption{The photonic lantern, a tapered waveguide that can adiabatically transfer light distributed between multiple fiber modes into multiple single-mode cores, or vice-versa. The particular lantern shown above is an idealization of a 3-port lantern, with all jacket, cladding, and core cross-sections assumed to be circular throughout the transition zone. Darker regions indicate higher refractive index. Adapted from \cite{Lin:21}. }
    \label{fig:PL}
\end{figure}
\\\\
In parallel, a number of new ideas and techniques are being proposed to further advance direct exoplanet characterization. One development is in short-exposure exoplanet imaging, which leverages statistical differences in planet and star speckle behavior at millisecond timescales to distinguish between planet light from starlight \cite{Rodack:21,Galicher:19}. This technique is distinct from ADI and SDI. Coherent detection, which exploits the incoherence of planet light, presents an alternative pathway for separating planet light and starlight. A related technique is nulling interferometry, an alternative to conventional coronagraphy that can achieve smaller inner working angles, and which works by destructively interfering starlight collected from different subapertures or telescopes. Other advances in direct characterization will need to be made not in the isolation of planet light, but the spectral analysis of that light. The high-resolution spectral analysis of faint objects like exoplanets will require methods for both the efficient coupling of light into the science instrument, and stabilization of that same light, which will vary with time due to passage through the atmosphere and instrument. These two requirements are typically in tension \cite{Lin:21}, and thus hard to achieve simultaneously.
\\\\
The photonic lantern (PL; \cite{Birks:15}) provides a capable platform for the above applications; other notable applications include OH line suppression through fiber Bragg gratings \cite{Trinh:13:GNOSIS,ellis:18:PRAXIS}, and spectroastrometry \cite{Gatkine:19}. As seen in Figure \ref{fig:PL}, the PL is a tapered waveguide that gradually transitions from a few-mode optical fiber (FMF) geometry to multiple widely-spaced single-mode cores, similar to a multi-core fiber (MCF), which can then be fanned out to an array of single-mode fibers (SMFs). When the FMF end is placed in the focal plane, the PL can efficiently couple multi-modal telescope light into multiple SMFs. While PLs come in a wide array of port counts and geometries, they can be largely classified into three groups. In what we call the ``standard'' PL, embedded cores are uniform in structure and refractive index. At the other extreme, ``mode-selective'' PLs use differing single-mode core radii or index contrasts, so that each fiber mode at the FMF-like lantern entrance routes to a distinct output port \cite{Leon-Saval:14}. Lastly, we term lanterns that operate between these two extremes ``hybrid lanterns.'' These lanterns have one core mismatched from the rest, thereby funnelling light from the fundamental fiber mode into a single output port while mixing the remaining light in the rest of the ports. This concept is similar to the ``mode-group selective'' lantern, introduced in \cite{Vel:18}. 
\\\\
Critically, in the process of coupling light into an array of SMFs, PLs map phase aberrations into intensity variations in a one-to-one manner, at least for small aberrations. This behavior enables the PL to additionally act as a 100\% duty cycle focal-plane WFS \cite{Corrigan:18,Norris:20,Wright:22}. Because PLs have a limited number of outputs (set by the manufacturing process, though PLs with up to 511 modes have been reported \cite{Birks:15}), these devices as of now can only give low-order wavefront information. Therefore, while PLs are well-suited to sense low-order aberrations like NCPAs \cite{Sauvage:07} and island modes \cite{NDiaye:18}, they are not a standalone WFS solution in XAO systems, which correct upwards of 1000 modes. In such applications, PLs will likely need to work in tandem with pupil-plane sensors like the Shack-Hartmann or pyramid WFS.
\\\\
We show an example of this phase-to-intensity mapping in Figure \ref{fig:astig}, which plots the non-degenerate intensity responses of a 6-port PL in the presence of positive and negative astigmatism. The focus of this work is to assess the performance of the photonic lantern wavefront sensor (PLWFS), in contexts like instrument coupling or coherent detection where PLs are already being considered for use. In these scenarios, the utility of the PL is doubled, enabling both the aforementioned non-WFS applications as well as focal-plane wavefront sensing. We focus on two contexts the first being fiber-fed, high-resolution spectrometry, mentioned above; and vortex-fiber nulling (VFN), a high contrast imaging technique which exploits symmetries in optical fiber modes to separate star and planet light \cite{Ruane:19:VFN}. In turn, we restrict our analysis to the infrared, since this wavelength regime will be the staging ground for the next push in direct exoplanet spectrometry, with upcoming instruments such as HISPEC and MODHIS \cite{Mawet:19:HISPEC}. 
\\\\
Research in PL wavefront sensing is ongoing. For instance, \cite{Norris:20} recently combined a 19-port PL with a neural net to enable nonlinear wavefront reconstruction of the first 9 non-piston Zernike modes. In comparison, we take a broader, but less in-depth approach: our goal is to provide a general baseline overview of the capabilities of the PLWFS, as well as the methods through which the sensing properties of these devices might be controlled. We place added emphasis on the linear analysis of the PLWFS, in order assess the limits of the PLWFS under more standard and simplistic linear AO control schemes. In Section \S\ref{sec:analytic}, we establish the math that will enable wavefront reconstruction with the PLWFS.
To begin, we present power series expansions for the PLWFS intensity response to first and second order in phase (\S\ref{ssec:genmodel}-\S\ref{ssec:second}). We also consider methods through which these models can be inverted, thereby enabling wavefront sensing. Next, we expand our models to arbitrary modal basis (\S\ref{ssec:modalbasis}): this both increases computational efficiency of the reconstruction models and allows them to be expressed in terms of common phase aberration bases such as the Zernike polynomials. In Section \S\ref{sec:analytic2}, we apply our models to quantify the behavior of the PLWFS. This analysis includes deriving conditions for WFS linearity (\S\ref{ssec:modeselectivity}-\S\ref{ssec:lincond}), and estimating maximum amount of WFE that can be handled by these sensors (\S\ref{ssec:range}).
\\\\
Finally, we combine our models with numerical simulations, to provide a first look at the wavefront-sensing abilities of a standard, hybrid, and mode-selective 6-port PL. Our aim in this work is to develop an initial understanding of the capabilities of the PLWFS, and in doing so we assume ``perfect'' lanterns and neglect noise (though we provide some reference to noise propagation in the linear regime in \S\ref{ssec:linearize}). We present an overview of our numerical method in \S\ref{sec:method}, and the corresponding results in \S\ref{sec:results}. In a companion paper \cite{paper2}, we extend these simulations to cover a range of PLWFS configurations beyond the 6-port geometries considered in this paper, in order to establish a rough baseline of the sensing abilities of PLWFSs. There, we also investigate potential strategies through which PLWFS performance can be further controlled and optimized. 

\section{Propagation analysis and phase reconstruction}\label{sec:analytic}
\subsection{General model}\label{ssec:genmodel}
\begin{figure}
    \centering
    \includegraphics[width=\textwidth]{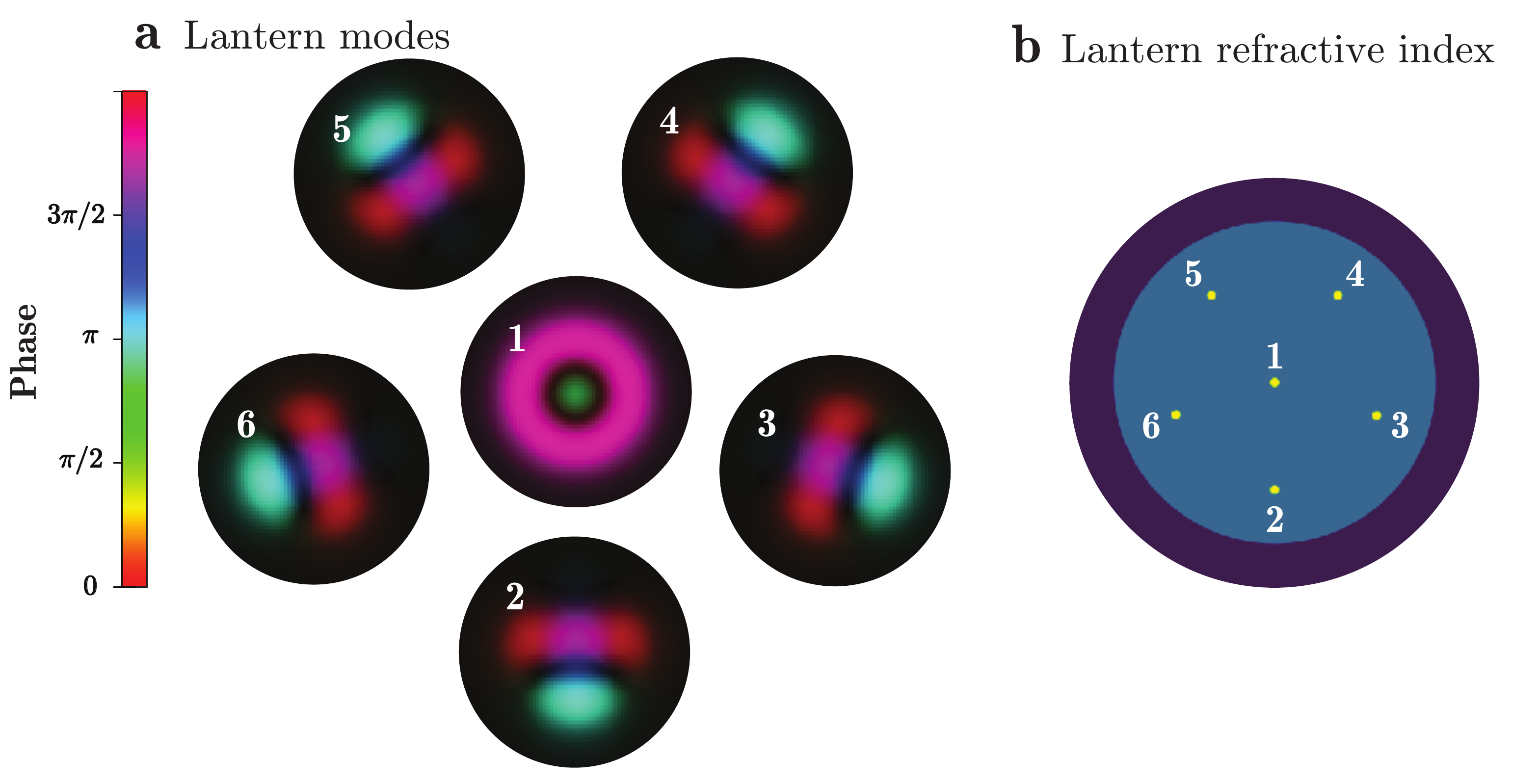}
    \caption{Panel a: Lantern modes for the same 6-port lantern as in Figure \ref{fig:astig}, evaluated at the lantern entrance. Phase is plotted in color, while amplitude is plotted in opacity. The 6 lantern modes are oriented to reflect the location of their corresponding lantern ports, shown on the right. To identify the ports and lantern modes, we index them according to the numerical labels. Panel b: the refractive index profile of the output (MCF-like) end of the PL. Embedded SMF cores are shown in yellow. Numerical labels connect each core to its corresponding lantern mode in panel a.  }
    \label{fig:lanternmodes}
\end{figure}

Consider the following general setup for a backend device to an AO-equipped telescope. AO-corrected light passes into an instrument backend, which may contain components such as beam-shaping (PIAA) optics \cite{Guyon:03:PIAA} and additional phase and/or amplitude optics (e.g. vortex fiber nuller mask). After light passes through some number of upstream components, it is focused onto the FMF end of a PL, ultimately propagating into the SMF ports at the PL output. These output ports may also optionally be inteferometrically combined. Because optical propagation is linear in complex electric field, the action of all backend optical components can be lumped into a single complex-valued transfer matrix, which we denote $A$. This matrix connects the input electric field ${\bm u}_{\rm in}$ and the output electric field ${\bm u}_{\rm out}$ of the backend device:
\begin{equation}\label{eq:amplitude}
    {\bm u}_{\rm out} = A {\bm u}_{\rm in}.
\end{equation}
In the case of the PLWFS, the transfer matrix $A$ will contain a projection component, since an $N$-port lantern will support only $N$ complex-valued electric field modes, meaning that the vector ${\bm u}_{\rm out}$ is $N$-dimensional. Note that, unlike the modes of a standard optical fiber, the modes of a PL are three-dimensional, encompassing the full propagation of light from the FMF-like input to the MCF-like output of the lantern. Here, we have a choice of mode basis. The modes we use in this work, which we term ``lantern modes,'' look like individual SMF modes at the lantern exit, and complex linear combinations of fiber modes at the lantern entrance. These modes can be computed by illuminating a single output core at the lantern exit and numerically back-propagating light to the lantern entrance.  Simulated cross-sections of lantern modes at the PL entrance, computed in this manner, are shown in Figure \ref{fig:lanternmodes} for a standard 6-port lantern. The $A$ matrix accounts for optical propagation through the telescope and any subsequent beam-shaping to the PL entrance, and then projects the focal plane electric field onto these lantern modes. Accordingly, $A$ has dimensions $N\times M$, for an $N$-port lantern and $M$ pupil samples.
\\\\
Since we ultimately measure intensity, not complex amplitude, we recast equation \ref{eq:amplitude} in terms of the intensity response $\bm{p}_{\rm out}$:
\begin{equation}\label{eq:intensity}
    \bm{p}_\text{out} = |A\bm{u}_\text{in}|^2.
\end{equation}
For phase-only aberrations, the goal of wavefront sensing is to invert equation \ref{eq:intensity} and recover the phase of $\bm{u}_{\rm in}$. We go over methods to do so in the following subsections.
\subsection{Linearizing intensity response}\label{ssec:linearize}
In this subsection, we provide a review of wavefront sensing in the linear regime. While optical propagation is linear in complex amplitude, it is nonlinear in intensity. However, for small changes in aberration amplitude, the intensity response will vary in a near-linear manner. Consider a phase-only aberration $\bm{\phi}$ in an electric field with assumed uniform intensity $I_{\rm in}=1$. We can approximate the intensity response of the system about some arbitrary reference phase $\bm{\phi}_0$ as
\begin{equation}\label{eqn:linstart}
    {\bm u}_{\rm in} = \exp(i\bm{\phi}) \approx  e^{i\bm{\phi}_0} \odot \left[\bm{1} + i (\bm{\phi}-\bm{\phi}_0)\right]
\end{equation}
where the vector $\bm{1}$ represents the electric field of a flat wavefront, and ($\odot$) represents element-wise (Hadamard) vector-vector multiplication. For clarity, we denote $\bm{\Delta \phi}\equiv \bm{\phi} - \bm{\phi}_0$, and modify the transfer matrix as $A_{ij} \rightarrow A_{ij}e^{i\phi_{0,j}}$; for a flat reference wavefront, $\phi_{0,j}=0$ and $A_{ij}$ is unchanged. The intensity resulting from the phase aberration $\bm{\phi}$ is
\begin{equation}
\begin{split}
    \bm{p}_\text{out} &= |A\bm{u}_\text{in}|^2 \\
    &\approx \big|A \left[\bm{1} + i \bm{\Delta\phi}\right]\big|^2\\
    &\approx |A\bm{1}|^2 + 2\,{\rm Im}\left[(A\bm{1})\odot (A^*\bm{\Delta \phi})\right]\\
\end{split}
\end{equation}
where the squaring and ($||$) operators are element-wise, and Im denotes taking the imaginary part. We can define the matrix $B$, having the same dimensions as $A$, as
\begin{equation}
    B_{ij} \equiv2\,{\rm Im}\left[ A^*_{ij} \sum_kA_{ik}\right]
\end{equation}
and recover 
\begin{equation} \label{eq:linear}
\bm{p}_\text{out} \approx |A \bm{1}|^2+B\bm{\Delta\phi}.
\end{equation}
We see that the first quantity represents the bias intensity when there is no phase error, while the matrix $B$ (often called the ``interaction matrix'' in the context of adaptive optics) describes the linear response of the intensities to phase perturbations from the reference wave -- in other words, $B$ is the Jacobian of the PL's intensity response, evaluated at the reference wavefront determined by $\bm{\phi}_0$. Equation \ref{eq:linear} can be inverted (e.g. via Moore-Penrose pseudo-inverse), enabling the reconstruction of phase errors from intensity responses. The phase aberration modes which this backend device can sense in the linear regime will be determined by the $B$ matrix; un-sensed aberration modes will lie in the null space of $B$. Alternatively, $A$ and $B$ can be used to compute gradients and Hessians for cost functions, enabling iterative nonlinear estimation for phase aberrations (see \cite{Frazin:18} for an example of this sort of analysis, with the pyramid WFS).
\\\\
Finally, to understand how error and noise propagates through the linear reconstruction process, we follow the analysis of \cite{Frazin:18}. We write the intensity response of an $N$-port PLWFS as
\begin{equation}
    \bm{p}_\text{out} = |A \bm{1}|^2 +B\bm{\Delta\phi} + \bm{n}(\bm{\Delta \phi}) +\bm{\nu}
\end{equation}
where $\bm{n}$ is an $N$-length vector of functions that accounts for the error caused by linearization, and $\bm{\nu}$ is the noise, assumed to be composed of $N$ independent zero-mean random processes. The least-squares estimate $\widetilde{\bm{\Delta\phi}}$ for the original phase aberration $\bm{\Delta\phi}$ is obtained through $B^+$, the pseudo-inverse of $B$, as follows:
\begin{equation}
\begin{split}
    \widetilde{\bm{\Delta\phi}} &= B^+ \left[\bm{p}_\text{out} - |A \bm{1}|^2\right]\\
    &= B^+B\bm{\Delta \phi} + B^+\left[\bm{n}(\bm{\Delta \phi}) +\bm{\nu}\right].
\end{split}
\end{equation}
Small singular values of $B$ will amplify both the error incurred by linearization, as well as random noise. Such amplification can be partially mitigated through regularization of the singular values.  Future analysis of the reconstruction properties of the PLWFS, particularly for nonlinear reconstruction and closed-loop operation, will require more detailed considerations of noise and error propagation --- we leave this for later work.  

\subsection{Second-order analysis of intensity response}\label{ssec:second}
Under perfect knowledge of the system transfer matrix $A$, we may obtain greater accuracy by expanding WFS response to second order. Express the incident electric field as 
\begin{equation}
        \bm{u}_{\rm in} \approx 
        e^{i\bm{\phi}_0}\left( 1 + i \bm{\Delta\phi} -\dfrac{1}{2}\bm{\Delta\phi}^2\right).
\end{equation}
Repeating the analysis of the previous subsection, again making the substitution $A_{ij} \rightarrow A_{ij}e^{i\phi_{0,j}}$, leads to the following:
\begin{equation}
    \bm{p}_{\rm out}
= |A \bm{1}|^2 + 2\,{\rm Im}\left[(A\bm{1})\odot (A^*\bm{\Delta \phi})\right]    
- {\rm Re} \, \left[(A \bm{1})\odot (A^*\bm{\Delta\phi}^2) \right] + |A \bm{\Delta\phi}|^2.
\end{equation}
We define the matrix $C$ as 
\begin{equation}\label{eq:defC}
    C_{ij} \equiv 2\,{\rm Re}\left( A^*_{ij} \sum_k A_{ik}\right)
\end{equation}
where Re denotes taking the real part. This yields the following formula for how phase errors up to second order affect intensity:
\begin{equation}\label{eq:2ndorder}
\bm{p}_{\rm out} \approx |A\bm{1}|^2 + B \bm{\Delta\phi} - \dfrac{1}{2}C \bm{\Delta\phi}^2 + |A\bm{\Delta\phi}|^2.
\end{equation}
Inversion of equation \ref{eq:2ndorder} can be accomplished using iterative techniques like Landweber iteration, the Levenberg-Marquardt algorithm, or gradient descent. Such methods often benefit from knowledge of the Jacobian, which can be derived from equation \ref{eq:2ndorder}:
\begin{equation}\label{eq:2ndorderJac}
    J_{ij} = \frac{\partial I_{\text{out},i}}{\partial\Delta\phi_j} = B_{ij}+(|A_{ij}|^2-C_{ij})\Delta\phi_j + \sum_{k} A_{ij}A^*_{ik} \Delta\phi_k.
\end{equation}
Our reliance on numerical solving techniques begs the question: prior to inversion, why approximate the intensity response of the PLWFS at all? An alternative strategy is to numerically solve equation \ref{eq:intensity} directly. However, we note two benefits of making the initial approximation. First, doing so simplifies the inverse problem, which improves numerical stability and mitigates issues where the numerical solver becomes stuck in local minima (similar to the phenomenon observed by \cite{Frazin:18}, for the pyramid WFS). This issue is exacerbated as the nonlinearity of a PL increases. Second, as we will later see in Section \S\ref{ssec:modalbasis}, truncation of the power series enables the use of a modal basis for phase aberrations, which increases computational efficiency, especially for a low-order sensors like the PLWFS. As an added note, the preliminary analysis presented in this work may one day enable or accelerate non-iterative nonlinear reconstruction akin to neural net methods \cite{Norris:20}, as well as more detailed analytic or semi-analytic characterization of the PLWFS.
\\\\
However, it is important to emphasize that inversion of quadratic and higher-order models is more complicated than their linear counterpart, primarily because nonlinear models can admit multiple solutions, at least for large WFE. This multiplicity may be a fundamental property of the WFS system, or an artifact due to truncation of the power series. We briefly discuss how some of these issues may be mitigated in \S\ref{ssec:quadrecon}.
\\\\
A cubic expansion is presented in Appendix \ref{ap:cube}.
\subsection{Modal basis}\label{ssec:modalbasis}
The matrices $A$, $B$, and $C$ will each have $N$ by $M$ entries, where $N$ is the number of output ports and $M$ is the number of sample points in the pupil plane. This is computationally inefficient --- the number of sample points will almost always greatly exceed the number of lantern ports, making the above matrices unnecessarily large. It is more efficient to represent phase aberrations in terms of some modal basis (e.g. the Zernike modes, the Karhunen-Lo\`eve modes derived from second-order phase aberration statistics, or the singular vectors of the $A$ matrix, projected onto pupil phase). To do so, write the phase aberration displacement vector $\bm{\Delta\phi}$ as
\begin{equation}\label{eq:modalbasis}
    \bm{\Delta\phi} = R\bm{a},
\end{equation}
where $\bm{a}$ is the real-valued vector of modal coefficients and $R$ is the change-of-basis-matrix, whose columns correspond to the basis vectors. Defining $B'\equiv BR$, the linear model given by equation \ref{eq:linear} is easily extended to modal basis as follows:
\begin{equation} \label{eq:linear_modal}
    \bm{p}_\text{out} \approx |A \bm{1}|^2+B'\bm{a}.
\end{equation}
Extension of the quadratic model to modal basis is more involved. Inserting equation \ref{eq:modalbasis} into equation \ref{eq:2ndorder} results in the following:
\begin{equation} \label{eq:quad_modal}
    \bm{p}_{{\rm out},i} \approx |A \bm{1}|^2_i + \left(B'\bm{a}\right)_i - \dfrac{1}{2}\sum_{jk} C'_{ijk} a_j a_k + |A'\bm{a}|^2_i
\end{equation}
where the tensor $C'$ is defined as 
\begin{equation}
    C'_{imn} \equiv \sum_j C_{ij} R_{jm}R_{jn}
\end{equation}
and the $A' \equiv AR$.
Differentiating equation \ref{eq:quad_modal} yields the Jacobian, under the quadratic approximation, in terms of modal basis:
\begin{equation}\label{eq:quadjac}
    J'_{ij} = B'_{ij} + \sum_k \left({\rm Re}\left[ A'_{ij}A'^*_{ik}\right] - \dfrac{1}{2}C'_{ijk}\right) a_k + \left( |A'_{ij}|^2 -\dfrac{1}{2}C'_{ijj} \right)a_j.
\end{equation}

\section{PLWFS properties}\label{sec:analytic2}
In this section, we provide an initial analysis into the wavefront-sensing properties of the PLWFS. Denote $\bm{u}_{\rm in}$ and $\bm{u}_{\rm out}$ as the input electric field (located in the pupil-plane) and output electric field (located at the backend of lantern), respectively, of the overall telescope-PLWFS system. The number of PL outputs is $N$.  Following the analysis of the previous section, $\bm{u}_{\rm in}$ and $\bm{u}_{\rm out}$ are related by the complex-valued transfer matrix $A$. Additionally, assume that there is no flux loss during propagation through the PL. We expand the $A$ matrix as a product of constituent matrices $U$, $P$, and $F$ such that
\begin{equation}
    \bm{u}_{\rm out} = UPF\bm{u}_{\rm in}.
\end{equation}
Here, $F \propto -i \mathcal{F}$ is the Fraunhofer propagator, where $\mathcal{F}$ is the Fourier transform. The $P$ matrix determines how the electric field couples into the lantern entrance, which resembles an FMF. More specifically, $P$ projects $\bm{u}_{\rm out}$ onto the basis of the $N$ first guided fiber modes for an FMF matching the geometry of the lantern entrance. For this work, our fiber modes are assumed to be the linearly polarized/LP modes, relevant for weakly guiding, circular, step-index optical fibers, which implies that $P$ is real-valued. Lastly, $U$ is the unitary matrix representing propagation through the lantern. In other words, $U$ transforms a focal-plane electric field, expressed in terms of LP mode amplitudes, into a set of complex-valued SMF amplitudes. Let us further assume phase-only aberrations. Expanding the complex exponential $e^{i\bm{\phi}}$ with Euler's identity yields
\begin{equation}
    \bm{u}_{\rm in} =  \bm{t}\odot \cos{\bm{\phi}} + i \bm{t}\odot \sin\bm{\phi}
\end{equation}
where $\bm{t}$ is the real-valued transmission mask of the pupil. We now derive some results.

\subsection{Impact of perfect mode selectivity}\label{ssec:modeselectivity}
In this section, we show that for an even pupil transmission $\bm{t}$, a ``perfect" mode-selective lantern (i.e. one free of manufacturing imperfections, which can separate the LP modes with zero crosstalk) maps $\pm \bm{\phi}$ to the same intensity response. This symmetry in the intensity response makes wavefront reconstruction impossible, preventing mode-selective lanterns from performing effectively as wavefront sensors. First, note that for such a lantern, the propagation matrix $U$ is the identity matrix. Therefore, the complex response of the system for a positive and negative phase aberration is
\begin{equation}
\begin{split}
\bm{u}_{\rm out}(\pm\bm{\phi}) &= -iP\mathcal{F}\left[\bm{t}\odot \cos{\bm{\phi}} \pm i \bm{t}\odot\sin\bm{\phi}\right]\\
&= -iP\left[ \bm{a} \pm i \bm{b}\right]
\end{split}
\end{equation}
where we have defined  
\begin{equation}
\begin{split}
        \bm{a} &\equiv \mathcal{F} \left[ \bm{t} \odot \cos\bm{\phi}\right], \\
        \bm{b} &\equiv \mathcal{F}\left[\bm{t}\odot\sin\bm{\phi}\right].
\end{split}
\end{equation}
We now make use of the following properties of the Fourier transform:
\begin{enumerate}
    \item The Fourier transform of a real, even function is real and even.
    \item The Fourier transform of a real, odd function is imaginary and odd.
\end{enumerate}
First, consider $\bm{\phi}$ even. In this case, due to the Fourier transform properties, the real-ness of $\bm{\phi}$, and the symmetry properties of composite functions, both $\bm{a}$ and $\bm{b}$ are real and even. Therefore, the intensity response is
\begin{equation}
    \bm{p}_{\rm out}(\pm \bm{\phi}_{\rm even}) = |\bm{u}_{\rm out}(\pm\bm{\phi}_{\rm even})|^2 = (P\bm{a})^2 + (P\bm{b})^2. 
\end{equation}
For even phase aberrations, the intensity response of a mode-selective PLWFS is even. Next, consider odd phase aberrations. Repeating a similar analysis, we now find that while $\bm{a}$ is still real and even, $\bm{b}$ is now odd and imaginary. Therefore, 
\begin{equation}\label{eq:odd_response}
    \bm{p}_{\rm out}(\pm \bm{\phi}_{\rm odd}) = |\bm{u}_{\rm out}(\pm\bm{\phi}_{\rm odd})|^2 = (P\bm{a})^2 + (iP\bm{b})^2 \pm 2(P\bm{a})\odot (iP\bm{b}).
\end{equation}
While an even phase aberration produces a real and imaginary field component, an odd phase aberration produces two real field components that interfere with each other. Under certain circumstances, this interference can break sign ambiguity. However, for the PLWFS, the vectors $\bm{a}$ and $\bm{b}$ are ultimately projected by $P$ onto the LP mode basis: a basis of real-valued, even and odd electric field distributions. As a result, the last term in equation \ref{eq:odd_response} is always 0. This is because $\bm{a}$ is even, and only has non-zero overlap with even modes, while $\bm{b}$ is odd, and only has non-zero overlap with odd modes. Finally, since any field can be decomposed into an even and odd component, the intensity response of the mode-selective PLWFS is even for all $\bm{\phi}$, at least in the vicinity of the origin.
\\\\
As a corollary, the above implies that mode-selective lanterns have a linear response matrix $B=0$. 

\subsection{Non-mode-selectivity can break sign ambiguity}\label{ssec:nonselective}
For a non-mode-selective lantern, the matrix $U$ is not the identity matrix; the rows of the matrix $UP$ are the (complex-conjugated) lantern modes. We repeat the analysis from the prior section. The intensity response is
\begin{equation}
    \bm{p}_{\rm out}(\pm \bm{\phi}) = |\bm{u}_{\rm out}(\pm\bm{\phi})|^2 = |UP(\bm{a}\pm i\bm{b})|^2.
\end{equation}
From the above, we see sign ambiguity is broken. The matrix $U$ applies a ``rotation" to the vector $P\bm{a}+iP\bm{b}$. While this rotation preserves the overall norm of the vector, it alters the the modulus of the individual elements, and hence, the powers in the individual ports of the PLWFS.
\\\\
In other words, switching the sign of a phase aberration is equivalent to conjugating the complex response of the telescope. If we immediately measure the focal plane electric field in the LP mode basis, this conjugation cannot be detected. However, if we apply a unitary transformation (e.g. a PL) after this conjugation, and then measure, the conjugation can be detected.
\subsection{Conditions for linearity}\label{ssec:lincond}
In this section we derive criteria that the PLWFS must meet to  maximize linear sensitivity to a given mode. We will restrict ourselves to the second-order expansion of intensity response for the PLWFS, equation \ref{eq:2ndorder}.
\\\\
To maximize the linear response of the PLWFS for a particular aberration mode, denoted by unit vector $\hat{\bm{z}}_i$, we require that the linear term in equation \ref{eq:2ndorder} is maximized and the quadratic terms are minimized. We can encourage this behavior by demanding that the quantity
\begin{equation}
Q\equiv \left[(A\bm{1})\odot (A\hat{\bm{z}}_i)^*\right]
\end{equation}
is purely imaginary. Repeating the same expansion of $A$ from the prior subsections, we equivalently require that
\begin{equation}\label{eq:lincond}
    Q\equiv \left[UP\mathcal{F}\bm{1}\odot (UP\mathcal{F}\hat{\bm{z}}_i)^* \right]
\end{equation}
is purely imaginary. To connect with the analysis of \S\ref{ssec:linearize},
note that $B\hat{\bm{z}}_i = 2 \, {\rm Im}\, Q$. Ultimately, linearity imposes a phase restriction on $Q$: linear response is maximized when $Q$ is purely imaginary, and minimized when $Q$ is purely real. Note that this maximization only enforces that intensity response of the PLWFS is predominately linear in the vicinity of the reference wavefront $e^{i\bm{\phi}_0}$ ; this is not a maximization of linear range, although it is likely the first step in an analytically-informed optimization of the latter.
\\\\
Optimization for the above metric entails designing a PL such that its corresponding propagation matrix $U$ satisfies equation \ref{eq:lincond}. This is tricky, but can be simplified in certain cases. In Appendix \ref{ap:6port}, we simplify the above linearity condition for a standard 6-port lantern in the presence of defocus.

\subsection{WFS limitations}\label{ssec:range}
Even with a perfect nonlinear reconstruction model, wavefront sensing breaks down when two distinct phase aberrations can map to the same WFS response. These ``degenerate" aberrations are not a concern when the WFS is operating in the linear regime and the mapping of aberrations to sensor intensity responses is one-to-one, but become increasingly problematic as the amplitude of phase aberrations increases. A way to estimate when degenerate aberrations may become problematic is to find input phase aberrations for which a column of the Jacobian becomes zero-valued. This estimate may be conservative, as the response of the PL in this regime can still carry useful information about the input WFE for a subset of the sensed modes.  Mathematically we look for an aberration vector $\bm{a}_0$ (defined, for instance, in Zernike basis) such that
\begin{equation} \label{eq:jac_extrema}
    \dfrac{\partial {\bm p}_{\rm out}}{\partial a_j}\bigg|_{\bm{a}_{0}} = 0.
\end{equation}
To motivate this criterion, suppose we find some aberration $\bm{a}_0$ where the above criterion is fulfilled. In turn, the WFS response about $\bm{a}_0$, in the $a_j$ direction, may behave quadratically:
\begin{equation}\label{eq:degen}
    \bm{p}_{\rm out}(a_{0,k}+a_j) = \bm{p}_{\rm out}(a_{0,k}) + \dfrac{\bm{p}_{\rm out}''(a_{0,j})}{2}  a_j^2 + o\left(a_j^3\right).
\end{equation}
Here, $a_{0,j}$ is the $j_{\rm th}$ element of $\bm{a}_0$. We immediately see that for small $a_j$, aberrations $a_{0,j}\pm a_j$ map to the same intensity response. More widely separated pairs of degenerate aberrations may also occur around $\bm{a}_0$, although they most likely will not be positioned symmetrically about $a_{0,j}$. For an alternative perspective, consider the modal-basis representation of the Jacobian, which has dimensions $N$ rows by $M$ columns for $N$ lantern ports and $M$ aberration modes, with $N\geq M$. The zeroing of a column makes the Jacobian rank-deficient, implying that locally about $\bm{a}_0$, the mapping of phase aberrations to PL intensity outputs can no longer be injective. In other words, we are guaranteed scenarios where two or more distinct phase aberrations map to the same intensity response.  
\\\\
The norm (or total RMS WFE) of the smallest aberration vector $\bm{a}_0$ which satisfies \ref{eq:jac_extrema} sets the scale in phase aberration space beyond which degeneracy can occur. We term this scale the ``degenerate radius."  To actually compute the degenerate radius, we take a numerical approach: feeding a standard root-solving algorithm (e.g. Levenberg-Marquardt) a series of random initial guesses in the vicinity of the origin, repeatedly solving \ref{eq:jac_extrema}, and then taking the solution with the smallest norm from the returned set. In this approach, we require the full form of the Jacobian for the WFS, without any power series approximations. We derive the following form for the Jacobian:
\begin{equation}
    \dfrac{\partial (p_{{\rm out},i}/p_{\rm in})}{\partial a_k} = - 2 \, {\rm{ Im}} \left[ \sum_{j} A_{ij}e^{i \phi_j} R_{jk} (1-a_k)  \times \sum_{j'} A^*_{ij'}e^{-i\phi_{j'}}   \right].
\end{equation}
Here, $\bm{\phi} \equiv R \bm{a}$, similar to section \S\ref{ssec:modalbasis}, with the exception that we are no longer expanding about some reference phase $\bm{\phi}_0$. A rougher but simpler approximation for the degenerate radius can be made by expanding wavefront response only to second-order: essentially, we set equation \ref{eq:quadjac} equal to 0, for fixed aberration index $j$. This conveniently gives an ordinary matrix-vector equation which can be solved quickly and directly using the Moore-Penrose pseudo-inverse, giving exactly one solution $\bm{a}_{0}$ per aberration. However, this approach can be inaccurate if the WFS response contains little quadratic component.  
\\\\
Lastly, we consider the maximum number of modes that an $N$-port lantern can sense. In the linear model, it is clear that such a lantern at most can sense $N$ aberration modes. However, this limit holds for nonlinear models as well. This is because our optical system, while nonlinear in intensity, is linear in complex amplitude. A lantern attempting to sense more aberration modes than it has ports is guaranteed to map two distinct phase aberrations to the same complex-valued lantern response, and in turn, the same real-valued intensity. Topological theorems, such as invariance of domain, lead to the same conclusion.

\section{Simulations}\label{sec:method}
In order to provide the initial steps for general characterization of the PLWFS, we simulate these devices using a numerical model in Python. This model has three primary components: a telescope model, which takes in an incident wavefront and returns a focal plane electric field; a PL propagator, which takes both a focal plane electric field and a lantern geometry, and returns the resulting power distribution of the output ports; and wavefront reconstructer, based on the analysis in Section \S\ref{sec:analytic}. Sections \S\ref{ssec:tele}, \S \ref{ssec:lant}, and \S\ref{ssec:recon} expand upon these components, respectively. Finally Section \S\ref{ssec:param} goes over the specific 6-port PL geometries which we simulate with our numerical model.

\subsection{Telescope simulation} \label{ssec:tele}
Propagation through telescope optics is handled using the HCIPy package \cite{hcipy}. Simulations are monochromatic, at a wavelength of 1.55 $\upmu$m. We additionally assume a 10 m circular, unobstructed aperture; the focal ratio of the system is optimized to ultimately maximize coupling of an unaberrated wavefront into the PL. Pupil-to-focal plane propagation is handled via HCIPy's Fraunhofer propagator.

\subsection{Lantern propagation} \label{ssec:lant}
After computing the focal-plane electric field distribution, the next step is to determine the corresponding electric field at the output of the lantern. To do so, we multiply the electric field vector by the lantern's propagation matrix, $UP$, which can be computed in pixel basis by discretizing the input plane of the PL and repeatedly propagating single-pixel electric fields. Alternatively, we can compute the lantern modes for a given PL design by illuminating each single-mode port at the lantern output with its fundamental mode and backpropagating light to the lantern entrance; the complex conjugate of the lantern modes form the rows of the propagation matrix. When the number of PL outputs is less than the number of pixels in the input plane, the backpropagation approach is more efficient; in this work, we use the latter. Numerical propagations through PLs are handled with the Lightbeam Python package \cite{mybpm}.

\subsection{Wavefront reconstruction}\label{ssec:recon}
Given some PLWFS intensity response, we may now attempt to reconstruct the original phase aberration. Critically, to simplify our models, we neglect the impact of noise; the treatment of noise, and related analyses of PLWFS sensitivity and closed-loop performance, are left for future work. In the meantime, our noiseless model will still be useful for an initial characterization of PLWFS capabilities. We also set our reference wavefront to be flat (i.e. in equation \ref{eqn:linstart} we set  $\bm{\phi}_0 = 0$)
Our reconstruction model is as follows.
\\\\
First, we expand phase aberrations in terms of the Zernike basis. To implement linear reconstruction, we compute the matrix $B'$ from equation \ref{eq:linear_modal}; this is done by numerically measuring the matrix of slopes $\partial I_i/\partial a_j$ about the origin. Here, $I_i$ denotes the intensity of the $i_{\rm th}$ output port and $a_j$ denotes the amplitude of the $j_{\rm th}$ Zernike mode, in radians RMS. We then calculate the Moore-Penrose pseudo-inverse of $B'$, which enables inversion of equation \ref{eq:linear_modal}. Note that this reconstruction method neglects any sort of flux normalization, which is unnecessary in the context of simulations but may be more desirable in a more practical implementation with real optics.
\\\\
In contrast, quadratic reconstruction requires knowledge of the $A$ matrix, equation \ref{eq:2ndorder}, which in turn determines the modal-basis matrices $A'$ and $B'$, and the tensor $C'$. The $A$ matrix can be computed by probing the pupil-plane electric field (resolved into a 128 by 128 grid of samples) pixel-by-pixel, and measuring the complex-valued response of the PLWFS, or alternatively through a backpropagation technique like in Section \S\ref{ssec:lant}. This is straightforward in the case of simulations, since the complex-valued electric field is known. In contrast, experimental measurement of the $A$ matrix will likely require some phase-diversity method. Inversion of the quadratic model, equation \ref{eq:quad_modal}, is handled using the Levenberg-Marquardt root-finding algorithm, as implemented by the Python package SciPy. We set the starting point of the root-finding routine to the linearly-reconstructed wavefront aberration.

\subsection{Simulated lanterns}\label{ssec:param}
To demonstrate the validity of our mathematical analysis, we simulate wavefront reconstruction with two types of 6-port PL: standard and hybrid. Both PLs obey the following assumptions. Firstly, we assume that PLs taper uniformly and linearly so that cross sections of the cores and overall cladding of a PL remain perfectly circular throughout the transition zone. While this is an idealization, it remains a useful starting point for a first-order analysis of the PLWFS, especially since it is unclear whether PL imperfections (such the non-circular claddings exhibited by PLs formed via the tapering of SMF bundles) will help or hurt sensing performance. 
\\\\
Beyond the above idealization, we assume that all PLs taper by a factor of 8 from entrance to exit, with cores spaced in the cladding in such a way that is consistent with the geometries produced when constructing lanterns from a bundle of uniformly sized SMFs. Cladding index is set to 1.444, corresponding to fused silica at 1.55 $\upmu$m wavelength, while jacket-cladding contrast is set to $5.5\times10^{-3}$; these parameters are typical for lantern construction (private communication with S. Leon-Saval). Core index is set so that the mode field diameter is $\sim$7.5 $\upmu$m, matching OFS ClearLite 980 16 fiber. The main difference between our simulated standard and hybrid PLs is in lantern core diameter. In the standard non-selective variant, all SMF cores have the same diameter ($4.4$ $\upmu$m), while in the hybrid variant one SMF core is made 2 $\upmu$m micron larger in diameter to accept the LP$_{01}$ mode. In either case, entrance diameter (i.e. the diameter of the cladding at the input FMF end of the lantern) is set to 20 $\upmu$m. Additionally, both lanterns have their lengths set by an optimization routine that maximizes for linearity in the lantern's intensity response to the first five non-piston Zernike aberrations. For more details on this procedure, see our companion paper \cite{Lin:21}.
\\\\
Lastly, as a sanity check, we also simulate a fully mode-selective variant of the 6-port lantern, to verify our result from \S\ref{ssec:modeselectivity} that such lanterns are insensitive to all aberration modes. For simplicity, we assume that the modes of this lantern are exactly the first 6 LP modes, bypassing the need for numerical beam propagation.

\section{Results}\label{sec:results}
In this section, we apply our numerical model to a standard, hybrid, and mode-selective 6-port lantern. In \S\ref{ssec:intensity}, we look at the intensity response of these PLs, in the presence of single aberrations. These response curves can be thought of as 1D slices of the PLWFS response ``surface'' in the presence of many aberration modes. Subsections \S\ref{ssec:linrecon} and \S\ref{ssec:quadrecon} compare the performances of the linear and quadratic reconstruction models in the presence of the first five non-piston Zernike aberrations. While this basis cannot fully describe ``realistic'' seeing conditions (and neglects any sort of cross-talk in the reconstruction process from higher order aberration modes) we leave analysis of low-order wavefront reconstruction in the presence of higher-order error for future work. Nevertheless, because the spatial-frequency spectrum of real WFE is typically very bottom-heavy \cite{Sauvage:07,NDiaye:18}, and because PLs are primarily sensitive to low-order modes, we believe that our simplified analysis is still useful.

\subsection{Intensity response}\label{ssec:intensity}
\begin{figure}
    \centering
    \includegraphics[width=\textwidth]{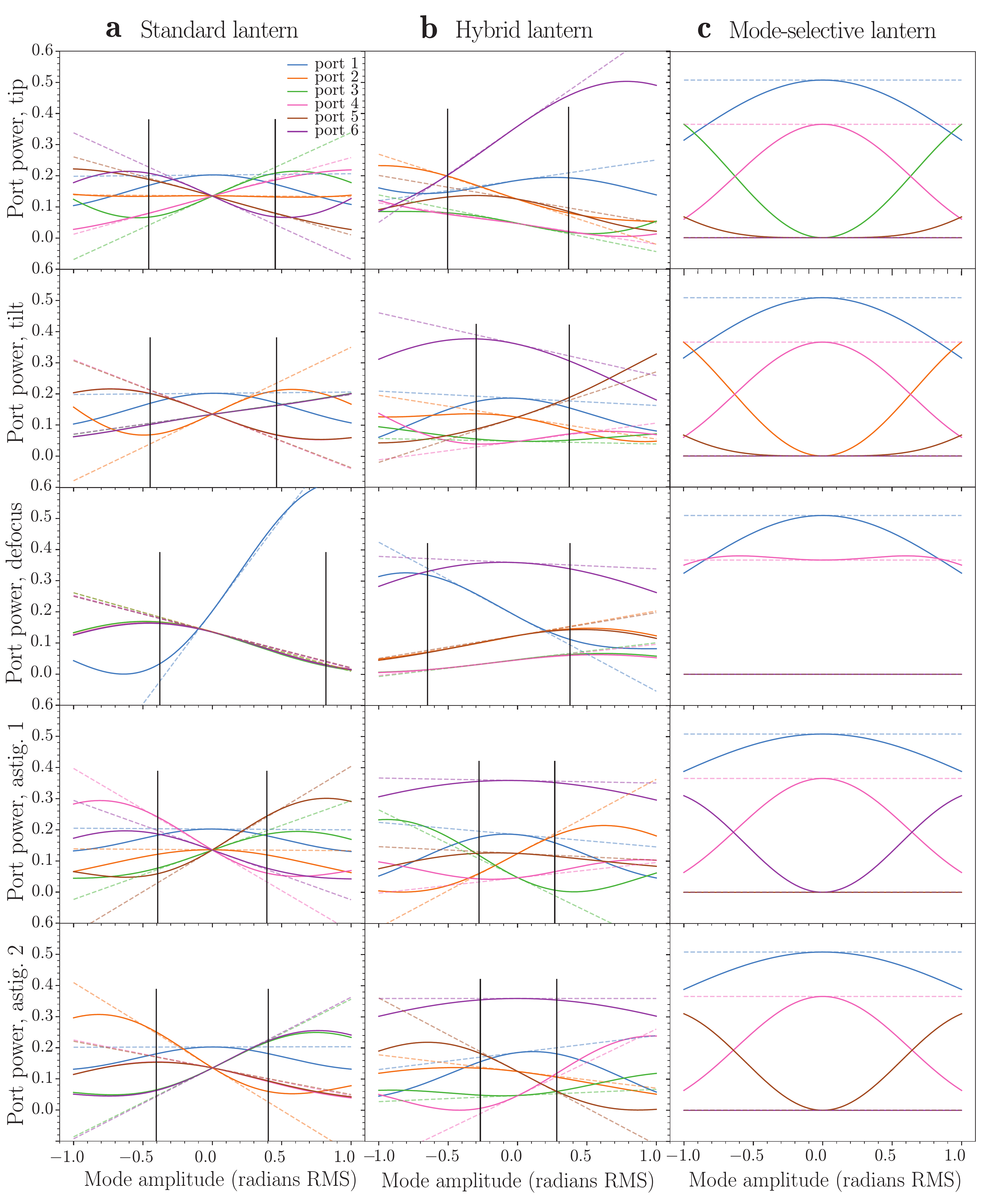}
    \caption{Column a: intensity response (solid, colored lines) of the 6 SMF outputs for a 6-port, standard lantern, as a function of aberration mode amplitude for Zernike modes 2-5 (tilt, defocus, and astigmatism). Vertical black lines denote the range where the linear model reconstructs the original aberration within 0.1 radians RMS. Dashed lines show the linear approximation for each port's response. Columns b, c: same as column a, but for a hybrid and mode-selective 6-port lantern, respectively.}
    \label{fig:intensity}
\end{figure}

Figure \ref{fig:intensity}a shows the intensity response of a standard 6-port lantern as a function of mode amplitude for the first 5 (non-piston) Zernike modes. Empirically, we find that this is the maximum number of modes a 6-port lantern can sense. In \cite{paper2}, we find the more general result that an $N$-port PL can sense at most $N-1$ Zernikes, without additional optics. Our heuristic explanation is that the complex-valued response of an $N$-port PL is sensitive to piston, which takes up one degree of freedom out of the $N$ total degrees in the system. This piston sensitivity is typically useless for wavefront sensing, and is lost in the conversion of complex amplitude to intensity. 
\\\\
We additionally mark the regions where the linear approximation holds. This ``linear range'' is defined as the interval in Zernike mode amplitude within which the linear model reconstructs the original phase aberration with less than 0.1 radians RMS of error. Intensity responses to the tilt and astigmatism modes exhibit good linearity in the interval around $[-0.4,+0.4]$ radians, while defocus exhibits linearity over a larger but more asymmetric range: around $[-0.4,0.8]$ radians. Note that the large linear range for defocus is primarily due to the taper length optimization outlined in \S\ref{ssec:param}. Conversely, certain values of taper length can lead to a lantern that is almost completely insensitive to defocus. We consider this and similar effects in more detail in our companion paper \cite{Lin:21}.
\\\\
Figure \ref{fig:intensity}b shows intensity responses for a 6-port hybrid lantern against the same modes. The introduction of a single, larger lantern core changes the lantern mode structure, both by replacing one of the modes with the LP$_{01}$ mode and by breaking the rotational symmetry of the lantern. We find that the 6-port hybrid lantern begins to behave nonlinearly more quickly than its non-selective counterpart. Additionally, as seen in Figure~\ref{fig:intensity}c, a fully mode-selective 6-port lantern has completely symmetric intensity responses, and therefore is not useful for wavefront sensing. This simulated result corroborates our analytic result from Section \S\ref{ssec:modeselectivity}. Finally, we find that our tested 6-port hybrid lanterns outperformed its standard counterpart in terms of degenerate radius (1.3 vs. 0.86 radians). This suggests that hybrid lanterns may exceed standard lanterns when using nonlinear reconstruction methods. 
\\\\
Crucially, we emphasize that the above results are for a specific subset of 6-port lantern geometries. In \cite{paper2}, we extend these results to a wider range of PL designs.

\subsection{Linear reconstruction}\label{ssec:linrecon}
\begin{figure}
    \centering
    \includegraphics[width=\textwidth]{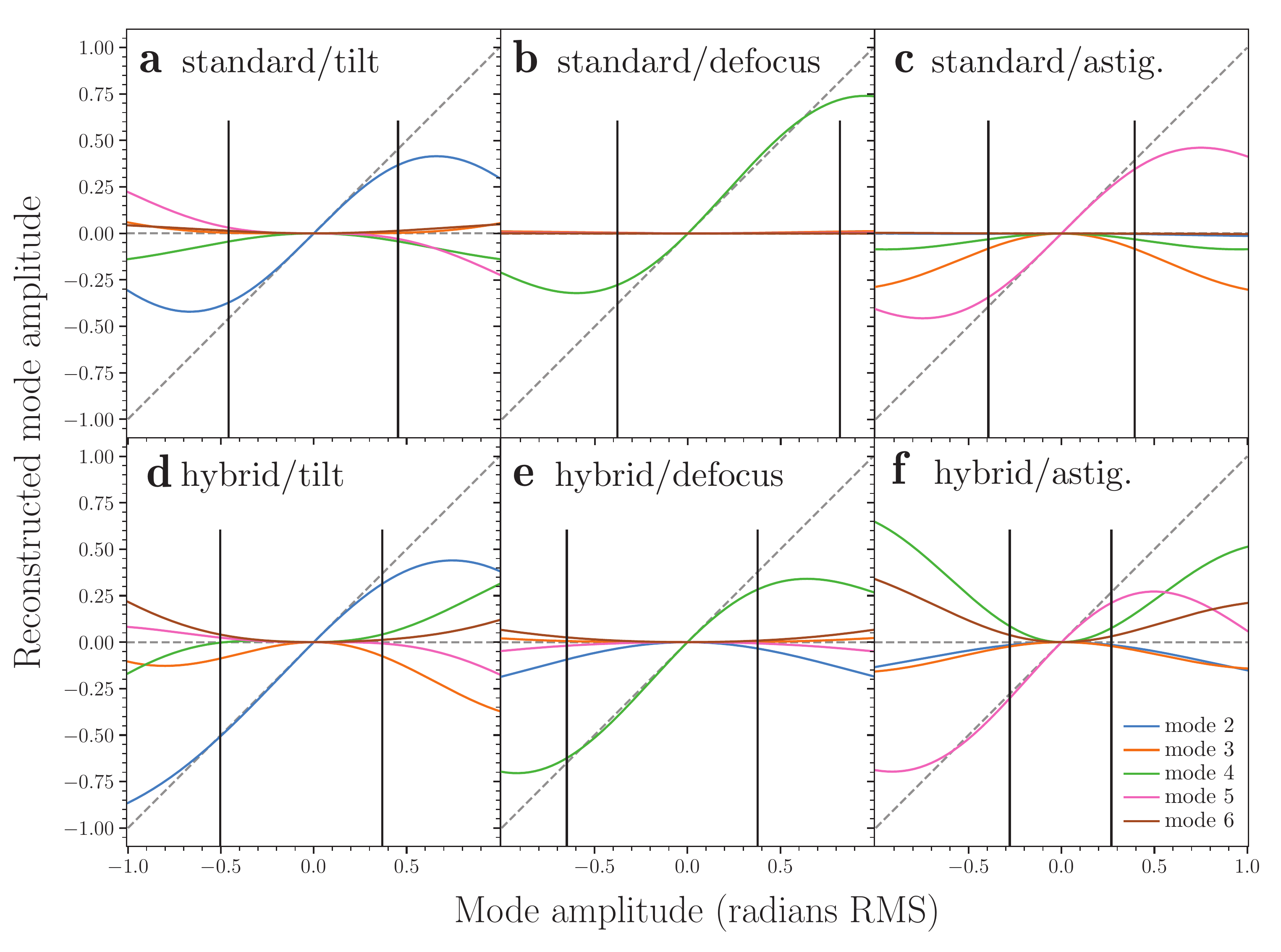}
    \caption{Panels a,b,c: Aberration reconstruction with the linear model, for various Zernike modes. Aberration amplitude is varied along the horizontal axis; this aberration is propagated into a lantern intensity response through our numerical model. We then attempt to do a linear reconstruction of the aberration amplitude, which we plot along the vertical axis. Under perfect reconstruction, the trace corresponding to the scanned mode should follow the line $y=x$ (marked by the diagonal, dashed gray line) while all other traces should follow $y=0$ (flat, dashed gray line). As before, vertical black lines mark the region where linear reconstruction is accurate within 0.1 radians RMS. Panels d,e,f: the same as previous panels, but for a 6-port hybrid lantern.  }
    \label{fig:lin_recon}
\end{figure}
Given the intensity responses in Figures \ref{fig:intensity}, computed over a range of Zernike mode amplitudes, we now apply the linear model (equation \ref{eq:linear_modal}) in an attempt to reconstruct the original mode amplitude. Figure \ref{fig:lin_recon} plots reconstructed aberration mode amplitude against true mode amplitude for Zernike tilt, defocus, and astigmatism, both for a standard and hybrid 6-port lantern. From the Figure, we see that in terms of reconstruction range, the hybrid lantern performs worse than the standard lantern in all modes, particularly in astigmatism. This is in line with results from \S\ref{ssec:intensity}.
\\\\
In order to test reconstruction performance in the presence of multiple aberrations, we use a Monte-Carlo approach. We first randomly draw 10,000 aberrated wavefronts (each composed of a random linear combination of Zernike modes 2-6), then pass each wavefront through our PLWFS model to obtain the corresponding intensity response. Given the intensity response, we attempt linear reconstruction. The root-mean-square of the difference between the ``true" wavefront and the reconstructed wavefront gives an estimate of the overall accuracy of the reconstruction scheme. Figures \ref{fig:recon_all_modes}a and b plots this accuracy against total aberration, for the standard and hybrid lantern, respectively. From the Figure, we see that reconstruction accuracy for the standard lantern remains under 0.1 radians for wavefront aberrations with up to $\sim 0.35$ radians of total RMS WFE; the hybrid lantern remains similarly accurate up to a lesser $\sim 0.25$ radians of total RMS WFE. While this result --- that hybrid lanterns behave more nonlinearly than standard lanterns --- is specific to 6-port PLs, we find in \cite{paper2} that it also applies for PLs of other sizes.

\subsection{Quadratic reconstruction}\label{ssec:quadrecon}
\begin{figure}
    \centering
    \includegraphics[width=\textwidth]{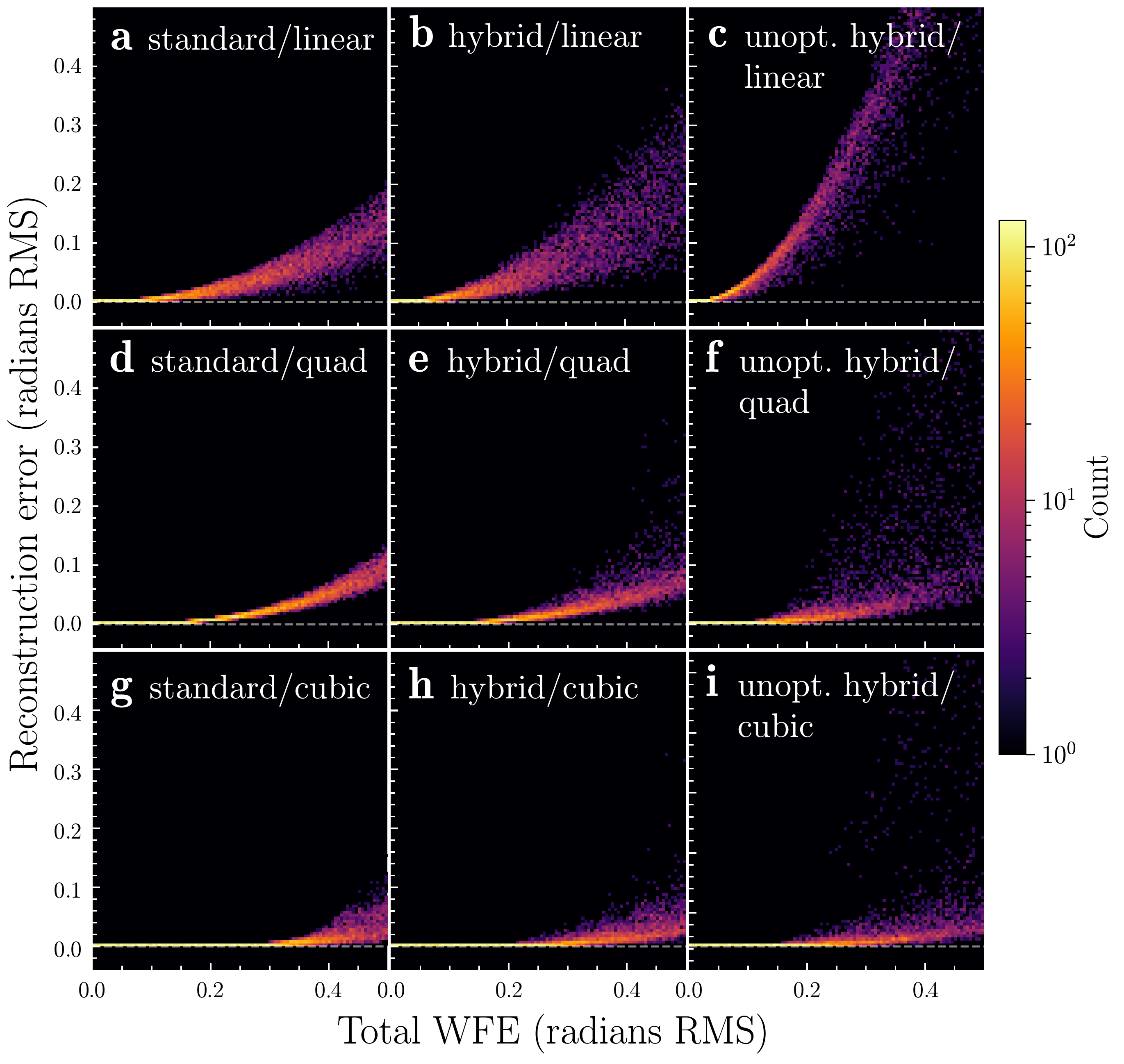}
    \caption{Panels a,d,g : heatmap of wavefront reconstruction accuracy against total RMS WFE for the standard 6-port lantern in the presence Zernike modes 2-5, using the linear, quadratic, and cubic reconstruction methods, respectively. Reconstruction accuracy was computed for 10,000 randomly sampled phase aberration vectors. A perfect wavefront sensor would have all reconstruction error along the line $y=0$, shown by the dashed gray line. Panels b,e,h : same as the previous panels but for the hybrid 6-port lantern. Panels c,f,i: same as the previous panels but for a hybrid 6-port lantern without linearity-maximizing taper length optimization. }
    \label{fig:recon_all_modes}
\end{figure}

In this subsection we present simulated results for the simplest nonlinear reconstruction method: quadratic reconstruction. This method is based off equation \ref{eq:quad_modal}, which we invert using the Levenberg-Marquardt root-finding algorithm as implemented by the Python package SciPy. For the initial guess required by the root-finder, we use the linearly reconstructed phase aberration vector. 
\\\\
We use the same Monte-Carlo approach outlined in the previous section to test the reconstruction performance of the quadratic model. Our results -- reconstruction accuracy against total RMS WFE for 10,000 randomly sampled aberrations -- are shown in Figure \ref{fig:recon_all_modes}d and e, for the standard and hybrid 6-port lanterns, respectively. Comparing with panels a and b, which were generated using the linear reconstruction model, we see that the quadratic model lowers the overall error in wavefront reconstruction, as expected. Specifically, for the standard lantern, quadratic reconstruction allows aberrations with up to $\sim 0.45$ radians of total RMS WFE to be reconstructed to an accuracy of 0.1 radians RMS. The hybrid lantern is similarly accurate up to $\sim 0.35$ radians of total RMS WFE. These results represents a $\sim 30-40$\% increase in reconstruction range over the linear model. Notably, the hybrid PL benefits more from quadratic reconstruction than the standard PL, which reinforces the notion that the hybrid PL behaves more nonlinearly.
\\\\
The quadratic model has the potential to provide even greater gains in reconstruction range when applied to PLs that behave more nonlinearly than the 6-port lanterns tested in this work, whose lengths were specifically optimized to maximize linearity. To show this, we apply the linear and quadratic reconstruction models to a 6-port hybrid lantern without any linearity optimization. Results are shown in Figures \ref{fig:recon_all_modes}c and f, respectively. The large spread and diverging pattern of points in panel c clearly shows the highly nonlinear nature of this particular PL; nevertheless, when switching to quadratic reconstruction model in panel f, the reconstruction error for most aberrations drops dramatically. In scenarios where linearity optimization is infeasible, quadratic reconstruction may provide an alternate path to improving WFS performance. 
\\\\
However, the quadratic model is not without downsides. The major issue is that quadratic reconstruction tends to become increasingly numerically unstable as total RMS WFE increases. We see this behavior reflected in Figure \ref{fig:recon_all_modes} particularly in panels e and f, where the scatter of points increases substantially with increasing RMS WFE. These instabilities can occur when the root-finder used to invert equation \ref{eq:quad_modal} gets stuck in a local minimum; a similar phenomenon was observed for the pyramid WFS in \cite{Frazin:18}). We discuss how this instability may be circumvented in Section \S\ref{sec:disc}.

\section{Discussion}\label{sec:disc}

In Section \S\ref{sec:analytic}, we laid out a general mathematical framework, in arbitrary modal basis, for the intensity response of a WFS to errors in phase. While we recover the usual linear model in our first-order expansion, we additionally derive a quadratic reconstruction model. This model can improve reconstruction accuracy, especially for PLs: the general nonlinearity of these devices often leads to quadratic-like intensity responses which are not well-fit by the linear model. However, the added accuracy of this scheme is offset by increased complexity: the inversion from intensity to aberration phase may require iterative methods that are slower than the linear model's single matrix multiplication. The higher-order nature of this model also introduced degeneracy, allowing for the mapping of two distinct phase aberrations to the same intensity response (though this is often simply a reflection of the fundamentally degenerate behavior of the PLWFS at large enough WFE). This degeneracy makes inversion more numerically unstable, and also enables scenarios where the root-solver becomes stuck in a local minimum. It remains to be seen whether the increased accuracy afforded by the quadratic and higher-order models outweigh the penalties in numerical stability and computation time, and if these techniques can be applied to closed-loop operation. We expect additional complications when moving to wavefront reconstruction with real PLWFSs. For one, we will have to contend with detector and photon noise, which will degrade both sensitivity and reconstruction range. Noise will likely be particularly problematic at the kHz refresh rates typically used for atmospheric compensation, but may be less of an issue when sensing slower NCPAs. An additional complication is that, in practice, the complex-valued $A$ matrix must be experimentally determined (e.g. through phase diversity methods), and hence will be prone to the effects of random and systematic uncertainties. While linear reconstruction, which requires only intensity knowledge, will be largely unaffected, uncertainties in $A$ may make nonlinear reconstruction even more numerically unstable. These uncertainties may be mitigated if we can constrain the $A$ matrix (for instance, through its modulus, or through $B$).
\\\\
We imagine several potential next steps in our mathematical analysis. One interesting continuation is to extend our phase-only aberration analysis to amplitude aberrations as well. Another is the expansion of WFS intensity response to third order, which we begin in Appendix \ref{ap:cube}; see also Figure \ref{fig:recon_all_modes}g-i, which plots the similar reconstruction heatmaps as panels a-f but for a cubic reconstruction model. Cubic expansion is particularly interesting because many PL intensity response functions (e.g. for the 6-port standard lantern, Figure \ref{fig:intensity}a) appear predominantly cubic. Figure \ref{fig:recon_all_modes} confirms that this expansion can offer a significant boost in reconstruction accuracy, especially for the 6-port standard lantern. However, the drawbacks are similar to the quadratic model. Each increase in order is accompanied by an increase in the model degeneracy, as well as an increase in the rank of the tensors required by the model.  
\\\\
More advanced reconstruction models may overcome these drawbacks. For one, stochastic optimization algorithms like simulated annealing, while computationally expensive, are one potential way to avoid local minima in the inversion process. Another idea is to use wavelength diversity, leveraging the chromatic dependence of the PLWFS response: extra measurements at multiple wavelengths may make the reconstruction process for our nonlinear models significantly easier. These measurements can be made though spectral dispersion of the PLWFS outputs, as in the so-called photonic ``TIGER'' configuration \cite{Saval:12:TIGER}. Lastly, we emphasize that while going to higher order may amplify numerical instability, it does not amplify experimental uncertainties in the $A$ matrix; this is because intensity will always have a second-order dependence on complex amplitude.
\\\\
Besides enabling wavefront reconstruction, mathematical models have a second, important use: they allow us to derive certain WFS properties and metrics through which the WFS can be optimized. For instance, in \S\ref{ssec:modeselectivity}, we derived that a fully mode-selective lantern is insensitive to phase aberrations, for even pupil transmission. It remains to be shown whether or not this limitation can be practically overcome with pupil masks or other additional optics. In contrast, there are no such restrictions with standard and hybrid lanterns. As a corollary, we found that the linearity of the PLWFS, at least for small aberrations, depends on the phase of what we call the $Q$ metric (equation \ref{eq:lincond}). We also show how this linearity condition simplifies for certain cases, such as the 6-port standard PL in the presence of defocus (Appendix \ref{ap:6port}). In the future, it may be desirable to optimize the PLWFS for this linearity condition. However, if nonlinear reconstruction methods, such as the quadratic or cubic methods in this work or the neural-net approach from \cite{Norris:20}, can be developed that are fast and stable enough to compete with linear reconstruction, it may instead be desirable to optimize lanterns according to degenerate radius (equation \ref{eq:degen}). Both the linear $Q$ metric and the degenerate radius are only the first steps in analytically defining the sensing properties of the PLWFS. Next steps will be to derive expressions for other potentially more useful properties, such as linear range (different from our condition \ref{eq:lincond}, which only ensures local linearity about the origin). Collectively, these analytically-derived expressions will help inform the manufacture of real PLWFSs in the future.
\\\\
Finally, we used our mathematical models to numerically simulate and compare the wavefront-sensing performance of an idealized standard, hybrid, and mode-selective 6-port PL. As expected, we recovered our analytic result that the mode-selective PL under even pupil illumination is insensitive to phase aberrations. We also found that the hybrid PL behaved more nonlinearly than the standard PL, suggesting that the latter may make a better wavefront sensor if used with a linear reconstruction scheme. In contrast, the larger degenerate radius of the 6-port hybrid lantern may make it a better choice with nonlinear reconstruction schemes. The next step will be to improve our model accuracy by accounting for manufacturing imperfections in simulated PLs, and to verify these models on an experimental testbed. 

\section{Conclusion}
In this work, we provide an end-to-end mathematical analysis of the PLWFS. In Sections \S\ref{sec:analytic} and \S\ref{sec:analytic2}, we developed linear and higher-order mathematical models for the intensity response of the PLWFS. These models enable the reconstruction of wavefront aberrations from intensity responses, and enable the derivation of certain metrics, such as the degenerate radius, which estimates the maximum amount of RMS WFE an aberration can have before the mapping of aberrations to intensities is no longer one-to-one. Such metrics can be used to benchmark and control the sensing behavior of these devices. Higher-order reconstruction models, such as quadratic (\S\ref{ssec:second} and \S\ref{ssec:quadrecon}) and cubic (Appendix \ref{ap:cube}), can additionally enable greatly improved reconstruction accuracy over the the standard linear model, but at the cost of added computation time and potentially increased numerically instability. Through our framework, we also show that a fully mode-selective lantern cannot sense wavefront aberrations with even pupil illuminations.   
\\\\
As a proof-of-concept, we apply our reconstruction models to a standard, hybrid, and mode-selective 6-port lantern in Section \S\ref{sec:results}, and successfully show that for the first two cases wavefront reconstruction of the first 5 non-piston Zernike modes is possible; 5 is the maximum number modes that can be sensed by either 6-port variant. We additionally confirm, numerically, that mode-selectivity (at least with an even pupil) hinders wavefront sensing. Comparing the performance of the standard and hybrid lanterns at a single output wavelength of 1.55 $\upmu$m, we find that the standard lantern has the highest linear range, accurately sensing the first five non-piston Zernike modes out to $\sim 0.5$ radians, followed by the hybrid lantern. Conversely, the 6-port hybrid PL outperformed the standard PL in terms of degenerate radius.
In the second part of this paper, we extend our analysis and simulate reconstruction performance for a range of PLs in various configurations. We additionally provide initial investigations into new strategies through which the sensing properties of PLs can be controlled and optimized. In the near future, we hope to verify our results with real, imperfect photonic lanterns, through experimental and on-sky testing, and in doing so, add to the next generation of focal-plane wavefront sensors.

%

\section*{Acknowledgements}

This material is based upon work supported by the National Science Foundation Graduate Research Fellowship Program under Grant No. DGE-2034835. Any opinions, findings, and conclusions or recommendations expressed in this material are those of the author(s) and do not necessarily reflect the views of the National
Science Foundation. This work was also supported by the National Science Foundation under Grant No. 2109232.

\begin{appendices}
\section{Defocus performance for standard 6-port lantern} \label{ap:6port}
The linearity criterion from \S\ref{ssec:lincond} can be simplified for a standard 6-port lantern, located in the focal plane of a telescope with a filled circular aperture, in the presence of defocus. We order our basis of LP modes as (LP$_{01}$, LP$_{02}$, rest of the LP modes) and our output ports as (central port, rest of the ports). For simplicity, we also assume a reference phase $\bm{\phi}_0=0$. Due to symmetry, both an unaberrated wavefront and a defocused wavefront will only couple into LP$_{01}$ and LP$_{02}$. Furthermore, the coupling coefficients will be real. Therefore, we can set 
\begin{equation}
    P\mathcal{F}\bm{1}\equiv 
    \begin{pmatrix}
    a \\
    b \\
    0 \\
    \vdots
    \end{pmatrix} ,
    P\mathcal{F}\bm{z}\equiv 
    \begin{pmatrix}
    d \\
    f \\
    0 \\
    \vdots
    \end{pmatrix} 
\end{equation}
where $a,b,c,d$ are real numbers and $\bm{z}$ is the vector corresponding to the defocus mode. Denoting the columns of the lantern propagation matrix $U$ as $\bm{c}_i$, we find that equation \ref{eq:lincond} becomes
\begin{equation}
    Q = ad|\bm{c}_1|^2 + af\bm{c}_1 \odot \bm{c}_2^* + bd \bm{c}_2\odot\bm{c}_1^* + bf|\bm{c}_2|^2.
\end{equation}
We want $Q$ to be ``as imaginary as possible.'' Clearly, the first and last terms are real, so a lantern that satisfies
\begin{equation}
    ad|\bm{c}_1|^2 + bf|\bm{c}_2|^2 =0
\end{equation}
will behave ``more linearly" than one that doesn't. The middle terms apply another condition: the each element in $\bm{c}_1$ should be $90^\circ$ out of phase with its corresponding element in $\bm{c}_2$. In turn, this condition implies that the LP$_{01}$ and LP$_{02}$ components for each lantern mode must be $90^\circ$ out of phase. We have verified this behavior numerically.
\\\\
It is also useful to consider the converse of the above conclusion. Suppose that the LP$_{01}$ and LP$_{02}$ mode coefficients are in phase. Then, $\bm{c}_1\odot\bm{c}_2^*$ will be real, and $Q$ will be purely real. Consequently, the linear $B$ matrix will be 0 - lantern response is locally quadratic.
\section{Cubic expansion}\label{ap:cube}
Expand the incident electric field with a phase $\bm{\phi}$ about a reference phase $\bm{\phi}_0$:
\begin{equation}
    \bm{u}_{\rm in} = e^{i\bm{\phi}_0} \odot \left[\bm{1} + i\bm{\Delta\phi} - \frac{1}{2}\bm{\Delta\phi}^2 - \frac{i}{6}\bm{\Delta\phi}^3 + o\left(\bm{\Delta\phi}^4\right)\right],
\end{equation}
where $\bm{\Delta\phi}\equiv \bm{\phi}-\bm{\phi}_0$. As before, the intensity response of the WFS is 
\begin{equation}
    \bm{p}_{\rm out} = \big| A \bm{u}_{\rm in}|^2
\end{equation}
where $A$ is the complex-valued transfer matrix of the overall optical system. Modifying $A_{ij} \rightarrow A_{ij}e^{i\phi_{0,j}}$ and combining the above two equations, keeping only terms up to third order, yields
\begin{equation}
    \bm{p}_{\rm out} \approx \bm{p}_{\rm out ,\, quad} - \frac{1}{3} {\rm Im}\left[ A\bm{1} \odot A^* \bm{\Delta\phi}^3 \right] + {\rm Im}\left[A\Delta\bm{\phi}\odot A^* \bm{\Delta\phi}^2\right]
\end{equation}
where $\bm{p}_{\rm out ,\, quad}$ is the quadratic approximation for output intensity, as per equation \ref{eq:2ndorder}. We now expand the above model to an arbitrary modal basis. Let $R$ be a change-of-basis matrix, such that $\bm{\Delta\phi}=R\bm{a}$. The additional terms from the cubic expansion can be expressed as a single tensor multiplication of the form 
\begin{equation}
    \sum_{lmn} D'_{ilmn} \, a_l a_m a_n
\end{equation}
where the tensor $D'_{ilmn}$ is defined as 
\begin{equation}
    D'_{ilmn} =  {\rm Im} \left[ -\frac{1}{3}\sum_j
    A_{ij}\sum_k A^*_{ik} R_{kl}R_{km}R_{kn} +\sum_{jk}A_{ij}A^*_{ik}R_{jl}R_{km}R_{kn}\right].
\end{equation}

\noindent The $D'$ tensor has dimensions $N\times M\times M\times M$ for an $N$-port lantern sensing $M$ aberration modes. The full cubic model, in modal basis, is 
\begin{equation} \label{eq:cubic_modal}
    \bm{p}_{{\rm out},i} \approx |A\bm{1}|^2_i + \left(B'\bm{a}\right)_i - \dfrac{1}{2}\sum_{jk} C'_{ijk} a_j a_k + |A'\bm{a}|^2_i + \sum_{lmn} D'_{ilmn}\, a_l a_m a_n
\end{equation}
Brief empirical testing with this model shows that it can provide a significant increase in reconstruction accuracy, especially for PLs that have already been optimized for linearity. Heatmaps of reconstruction error against total RMS WFE for 10,000 randomly sampled aberrations are shown in Figure \ref{fig:recon_all_modes}g, h, and i, for various 6-port lantern designs. Notably, going to higher order consistently extends the reconstruction range of the sensor, suggesting that the main downside of going to a higher-order model is additional computational complexity rather than numerical instability, at least for the first few orders.
\end{appendices}

\bibliography{refs}






\end{document}